\documentclass[12pt]{article}
\usepackage{amssymb}
\usepackage{amsfonts}

\def\beq{\begin{equation}}
\def\eeq{\end{equation}}
\def\bb{\begin{eqnarray}}
\def\ee{\end{eqnarray}}

\begin{document}
\begin{titlepage}

\begin{center}
{\Large\bf The content of $f(R)$ gravity}
\end{center}
\vspace{0.5cm}
\begin{center}

{\large Alvaro N\'u\~nez\footnote{e-mail address:
an313@scires.nyu.edu}  and Slava Solganik\footnote{e-mail address:
ss706@scires.nyu.edu}}
\vspace{0.5cm}

{\em New York University, Department of Physics, New York, NY 10003, USA}.\\
\end{center}

\hspace{1.0cm}

\begin{abstract}

We analyze the propagating degrees of freedom in gravity models where the scalar
curvature in the action is replaced by a generic function $f(R)$ of the curvature. That these gravity models are equivalent to
Einstein's gravity with an extra scalar field had previously been shown by applying a conformal transformation. We confirm this result
by calculating the particle propagators. This provides further evidence of the unability of these models to explain the accelerated expansion 
of the Universe without  contradicting solar system experiments.
\end{abstract}

\end{titlepage} 

\newpage 

\bigskip

A suitable modification of gravity at large distances could explain the current accelerated expansion of the Universe observed in measurements 
of type 1a supernovae \cite{Riess:1998cb}. This idea had been suggested in the work \cite{Deffayet:2001pu}, 
the authors of which developed it further in the context of the proposed in \cite{Dvali:2000hr} model.  
On the other hand, cosmologically motivated theories that explain the small acceleration rate of the Universe
via modifications of gravity at very large scales can be tested in solar system experiments \cite{Dvali:2002vf}, making these infrared modifications 
an even more attractive subject for studying.

In this article we address the question of what other than DGP \cite{Deffayet:2001pu}, \cite{Dvali:2000hr} kinds of Einstein's gravity modifications 
would make a patology free theory.
Bearing in mind the acceleration problem,  
we will concentrate on the attractively simple modifications of Einstein's equation through  changing the
action by replacing the scalar curvature with a function of it \cite{Carroll:2003wy}. Specifically, we start from a
sufficiently generic theory with the action 
\bb \label{action} S=\int d^4x\,\sqrt{-g}f(R) + S_{matter}, \ee 
where $f(R)$ is some function of the scalar curvature and $S_{matter}$ is the action for matter fields.
It was shown some time ago that, by use of a conformal transformation, these kind of theories reduce essentially to the theory of Einstein's 
gravity plus that of an extra scalar
field (see, for example, \cite{Barrow:xh}, \cite{Kalara:1990ar}). 
The decoupling of this scalar mode would make these theories no
different from quintessence. 
However, the decoupling of the scalar mode would violate general covariance thus leading to an
unsatisfactory result. 
Moreover, there exists an argument about the existence of a preferred frame, questioning thus the legitimacy of
such conformal transformation \cite{Faraoni:hp}. 
The presence of a coupled to matter scalar mode would mean that
we are dealing with a scalar tensor  theory, which would have a valid perturbation expansion
but an incorrect Newtonian limit.

Here we will consider the modification (\ref{action}) from the point of view of propagating degrees of freedom by deriving the particle 
propagators of this theory. The variation of  the action with respect to the metric leads to the equations of motion
\bb
\label{eq}
f'(R)R_{\mu\nu}-{1\over2}g_{\mu\nu}f(R)+(g_{\mu\nu}\nabla_\lambda\nabla^\lambda-\nabla_\mu\nabla_\nu)f'(R)=T_{\mu\nu}.
\ee
One can see that for $f(R)=R$ they give the standard Einstein equation. These equations allow a constant curvature 
$R=R_0={\rm const}$ solution, $R_{0}$ being defined by the following condition
\bb
f'(R_0)R_0=2f(R_{0}).
\ee
Since this  is a maximally symmetric solution, it implies
\bb
R_{\lambda\mu\nu\sigma}=\frac{R_0}{12}(g_{\lambda\nu}g_{\mu\sigma}-g_{\lambda\sigma}g_{\mu\nu})
\ee
and
\bb
R_{\mu\nu}=\frac{1}{4}R_0g_{\mu\nu}.
\ee

We linearize now the equations of motion on the constant curvature solution background. 
We take the metric in the form $g_{\mu\nu}=g_{\mu\nu}^{(0)}+h_{\mu\nu}$, where $g_{\mu\nu}^{(0)}$ is the solution of (\ref{eq}) 
corresponding to our constant scalar curvature $R_0$, next expand the equations of motion till the linear order terms. Inverting the 
operator acting on $h_{\mu\nu}$, we can find the propagator. In order to write down this propagator, it is convenient to use spin 
projectors $P^2, P^1_m, P^1_e, P^1_b, P^1_{me}, P^1_{em}, P^0_s, P^0_w, P^0_{sw}, P^0_{ws}$ 
\cite{VanNieuwenhuizen:fi} (the explicit form of the projectors is given in the Appendix):
\bb
&-& \frac{P^2}{(\nabla^2 +R/2)f'(R)}-  \frac{P^1_m}{(R/2) f'(R)}-\nonumber\\
&-& \frac{f'(R)/2 +R/4 f''(R)}{(\nabla^2 +R/2)f'(R)\left(f''(R)(3\nabla^2 +R)-f'(R)\right)}P^0_s-\nonumber\\
&-&\frac{ \sqrt{3} \left(f'(R)/2 -(\nabla^2+R/4) f''(R)\right)}{(\nabla^2 +R/2)f'(R)\left(f''(R)(3\nabla^2 +R)
-f'(R)\right)}\left(P^0_{sw}+P^0_{ws}\right) +\nonumber\\
&+&\frac{\left((4\nabla^2 +R/2)f'(R)-3(2\nabla^2+R/2)^2 f''(R)\right)}{(\nabla^2 +R/2)R f'(R)\left(f''(R)(3\nabla^2 +R)-f'(R)\right)}P^0_w. 
\ee
In particular, for the flat background Einstein equation ($f(R)\equiv R, R=0$), we get the standard result for the graviton propagator
\bb
- \frac{P^2}{\nabla^2}+\frac{1}{2}\frac{P^0_s}{\nabla^2},
\ee
where the seemed to be ghost $P^0_s$ is just a cancelation term for the mass degree of 
freedom of the massive spin-2 projector $P^2$.

To analyze the particle content of the theory, we have to look at the $P^2$ and $P^0_s$ terms. The corresponding part of the propagator 
can be rewritten as
\bb
- \frac{\left(P^2-\frac{1}{2}P^0_s\right)}{(\nabla^2 +R/2)f'(R)} -  \frac{P^0_s}{(\nabla^2+R/3-f'(R)/(3f''(R)))f'(R)},
\ee
thus we have a massless graviton in a curved background, $-P^2+1/2\;P^0_s$, and a scalar particle, which becomes a ghost only for $f'(R)<0$. 
This condition, however,
cannot be satisfied since it would make graviton itself a ghost. 
Our result agrees with that  obtained by the conformal transformation,
according to which the modification of gravity by use of some arbitrary enough function 
of the scalar curvature is equivalent to the addition of a scalar field with a specific type of potential.
We also would like to notice that the recently suggested model of 
ghost inflation \cite{Arkani-Hamed:2003uy} 
doesn't reduce to the above modification even ignoring higher derivatives terms.

The modified gravity model should give us a reasonably flat observable Universe, which implies a tiny mass for
the scalar field. However, light scalar gravity is in contradiction with  solar system experiments.
It seems that it is impossible to decouple the scalar mode even by fine tuning the function $f(R)$. As one can see, its derivative defines the 
coupling of both tensor and scalar modes,
the $f(R)$ model is essentially a scalar tensor gravity.
Thus we have to conclude that such modifications of gravity are ruled out.
On the other hand it seems that discussions of a preferred physical frame are irrelevant,
at least in the linear approximation. 

We would like to thank Gia Dvali for useful discussions.

\section*{Appendix}
Below we give the expressions for the ten operators which span the space of solutions to the linearized 
field equations \cite{VanNieuwenhuizen:fi}: 
\bb
P^2&=&\frac12(\theta_{\mu\rho}\theta_{\nu\sigma}+\theta_{\mu\sigma}\theta_{\nu\rho})-\frac13\theta_{\mu\nu}\theta_{\rho\sigma},
\nonumber\\
P^1_m&=&\frac12(\theta_{\mu\rho}\omega_{\nu\sigma}+\theta_{\mu\sigma}\omega_{\nu\rho}+\theta_{\nu\rho}\omega_{\mu\sigma}+\theta_{\nu\sigma}\omega_{\mu\rho}),
\nonumber\\
P^1_e&=&\frac12(\theta_{\mu\rho}\omega_{\nu\sigma}\theta_{\mu\sigma}\omega_{\nu\rho}-\theta_{\nu\rho}\omega_{\mu\sigma}+\theta_{\nu\sigma}\omega_{\mu\rho}),
\nonumber\\
P^1_b&=&\frac12(\theta_{\mu\rho}\theta_{\nu\sigma}-\theta_{\mu\sigma}\theta_{\nu\rho}),
\nonumber\\
P^1_{me}&=&\frac12(\theta_{\mu\rho}\omega_{\nu\sigma}-\theta_{\mu\sigma}\omega_{\nu\rho}+\theta_{\nu\rho}\omega_{\mu\sigma}-\theta_{\nu\sigma}\omega_{\mu\rho}),
\nonumber\\
P^1_{em}&=&\frac12(\theta_{\mu\rho}\omega_{\nu\sigma}+\theta_{\mu\sigma}\omega_{\nu\rho}-\theta_{\nu\rho}\omega_{\mu\sigma}-\theta_{\nu\sigma}\omega_{\mu\rho}),
\nonumber\\
P^0_s&=&\frac13\theta_{\mu\nu}\theta_{\rho\sigma},
\nonumber\\
P^0_w&=&\omega_{\mu\nu}\omega_{\rho\sigma},
\nonumber\\
P^0_{sw}&=&\frac1{\sqrt3}\theta_{\mu\nu}\omega_{\rho\sigma},
\nonumber\\
P^0_{ws}&=&\frac1{\sqrt3}\omega_{\mu\nu}\theta_{\rho\sigma},
\nonumber
\ee
where the transversal and longitudinal projectors in the momentum space are respectively
\bb
\theta_{\mu\nu}=\delta_{\mu\nu}-\frac{\nabla_\mu \nabla_\nu}{\nabla^2},\qquad \omega_{\mu\nu}=\frac{\nabla_\mu \nabla_\nu}{\nabla^2}.
\nonumber
\ee




\begin{thebibliography}{99}

\bibitem{Riess:1998cb}
A.~G.~Riess {\it et al.}  [Supernova Search Team Collaboration],
Astron.\ J.\  {\bf 116}, 1009 (1998)
[arXiv:astro-ph/9805201].
S.~Perlmutter {\it et al.}  [Supernova Cosmology Project Collaboration],
Astrophys.\ J.\  {\bf 517}, 565 (1999)
[arXiv:astro-ph/9812133].
J.~L.~Tonry {\it et al.},
Astrophys.\ J.\  {\bf 594}, 1 (2003)
[arXiv:astro-ph/0305008].

\bibitem{Deffayet:2001pu}
C.~Deffayet, G.~R.~Dvali and G.~Gabadadze,
Phys.\ Rev.\ D {\bf 65}, 044023 (2002)
[arXiv:astro-ph/0105068].

\bibitem{Dvali:2000hr}
G.~R.~Dvali, G.~Gabadadze and M.~Porrati,
Phys.\ Lett.\ B {\bf 485}, 208 (2000)
[arXiv:hep-th/0005016].

\bibitem{Dvali:2002vf}
G.~Dvali, A.~Gruzinov and M.~Zaldarriaga,
Phys.\ Rev.\ D {\bf 68}, 024012 (2003)
[arXiv:hep-ph/0212069].
A.~Lue and G.~Starkman,
Phys.\ Rev.\ D {\bf 67}, 064002 (2003)
[arXiv:astro-ph/0212083].
A.~Lue, R.~Scoccimarro and G.~Starkman,
Phys.\ Rev.\ D {\bf 69}, 044005 (2004)
[arXiv:astro-ph/0307034].
G.~Dvali,
arXiv:hep-th/0402130.

\bibitem{Carroll:2003wy}
S.~M.~Carroll, V.~Duvvuri, M.~Trodden and M.~S.~Turner,
arXiv:astro-ph/0306438.
S.~Nojiri and S.~D.~Odintsov,
arXiv:hep-th/0308176.
R.~Dick,
arXiv:gr-qc/0307052.
D.~N.~Vollick,
Phys.\ Rev.\ D {\bf 68}, 063510 (2003)
[arXiv:astro-ph/0306630].
S.~Nojiri and S.~D.~Odintsov,
Phys.\ Lett.\ B {\bf 576}, 5 (2003)
[arXiv:hep-th/0307071].
A.~D.~Dolgov and M.~Kawasaki,
Phys.\ Lett.\ B {\bf 573}, 1 (2003)
[arXiv:astro-ph/0307285].
E.~E.~Flanagan,
Class.\ Quant.\ Grav.\  {\bf 21}, 417 (2003)
[arXiv:gr-qc/0309015].
X.~H.~Meng and P.~Wang,
arXiv:hep-th/0309062.
X.~H.~Meng and P.~Wang,
arXiv:astro-ph/0308284.

\bibitem{Barrow:xh}
J.~D.~Barrow and S.~Cotsakis,
Phys.\ Lett.\ B {\bf 214} (1988) 515.

\bibitem{Kalara:1990ar}
S.~Kalara, N.~Kaloper and K.~A.~Olive,
Nucl.\ Phys.\ B {\bf 341}, 252 (1990).

\bibitem{Faraoni:hp}
V.~Faraoni and E.~Gunzig,
Int.\ J.\ Theor.\ Phys.\  {\bf 38}, 217 (1999)
[arXiv:astro-ph/9910176].
V.~Faraoni, E.~Gunzig and P.~Nardone,
Fund.\ Cosmic Phys.\  {\bf 20}, 121 (1999)
[arXiv:gr-qc/9811047].

\bibitem{VanNieuwenhuizen:fi}
P.~Van Nieuwenhuizen,
Nucl.\ Phys.\ B {\bf 60}, 478 (1973).
P.~Van Nieuwenhuizen,
Phys.\ Rept.\  {\bf 68}, 189 (1981).

\bibitem{Arkani-Hamed:2003uy}
N.~Arkani-Hamed, H.~C.~Cheng, M.~A.~Luty and S.~Mukohyama,
arXiv:hep-th/0312099.
N.~Arkani-Hamed, P.~Creminelli, S.~Mukohyama and M.~Zaldarriaga,
arXiv:hep-th/0312100.
 
\end{thebibliography}
\end{document}